# Do Japanese and Italian women live longer than women in Scandinavia?

Ørnulf Borgan, Department of Mathematics, University of Oslo, Norway

## Abstract

Life expectancies at birth are routinely computed from period life tables. Such period life expectancies may be distorted by selection when comparing countries where the living conditions improved earlier (like Norway and Sweden) with countries where they improved later (like Italy and Japan). One way to get a fair comparison between the countries, is to use cohort data and consider the expected number of years lost before a given age *a*. Contrary to the results based on period data, one then finds that Italian women may expect to lose more years than women in Norway and Sweden, while there are no indications that Japanese women will lose fewer years than Scandinavian women.

## 1. Introduction

Since the Second World War, one has seen a remarkable increase in life expectancy at birth in a number of countries. In this paper, we consider the expected life length for women in Italy, Japan, Norway and Sweden. Figure 1 shows how life expectancy at birth for these women has increased from 1950 and until 2012. In 1950, women in Norway and Sweden had the longest life expectancy among the four countries with 73.3 years for Norway and 72.4 years for Sweden, while the expected life length was 67.5 years for Italian women and 60.9 years for Japanese women. In 2012 (which is the last year with data for Italy and Japan in the Human Mortality Database), the situation was completely different. Now the life expectancy at birth for women in Japan and Italy were 86.4 years and 84.5 years, respectively; clearly ahead of Norway (83.4 years) and Sweden (83.5 years). Based on these numbers, it seems as if Japanese and Italian women may expect to live longer than women in Scandinavia, and this has e.g. been attributed to a healthy lifestyle and diet (e.g. Willett, 1994).

But the situation is more complicated than it may seem at first look. The life expectancies shown in Figure 1 are computed from period life tables, and they may be interpreted as the expected life length of a newborn girl *if the age-specific period mortalities would continue into the future*. But this is a hypothetical situation, and real women do not live in this way. They are members of a birth cohort in a given country and live their lives under the changing living conditions of that country (disregarding migration for the sake of the argument). Not all the women in a cohort are equally robust, however, and the frail ones tend to die first leaving a selected population of more robust women at the higher ages (Vaupel et al. 1979). Moreover, when the living conditions in a country improve, the selection pressure becomes smaller and more of the frail women will live longer. Thus for a country where the living conditions improved earlier (as was the case in Norway and Sweden) one will expect the elderly population in our days to contain more frail women than it does in countries where the living conditions improved later (as they did in Japan and Italy). This makes it difficult to compare period life expectancies for different countries when they have had very different histories regarding the development of their living conditions (Aalen et al. 2008, section 6.5.3).

**Figure 1:** *Life expectancy at birth for females in Italy, Japan, Norway and Sweden for the periods from 1950 to 2012. (Source: The Human Mortality Database.)*

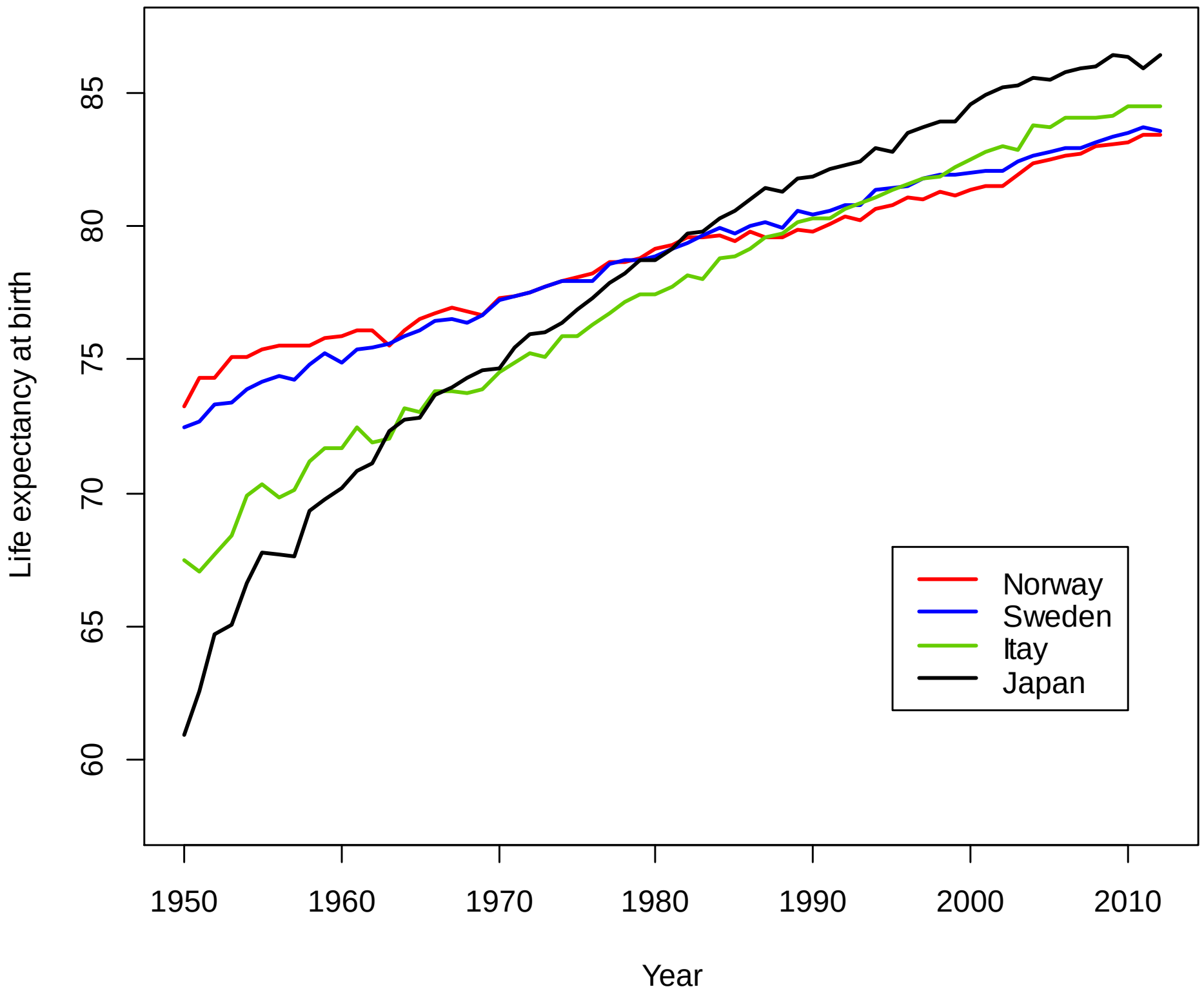

To obtain a fair comparison of how long women in different countries may expect to live, we should therefore consider cohort data, not period data. But then we are faced with a new problem. In order to compute the cohort life expectancy, we need to know the complete mortality history of a cohort, and this is only possible for cohorts that were born more than one hundred years ago. For younger cohorts, we are not able to compute the cohort life expectancy. A solution to the problem is not to consider life expectancy at birth, but rather use the *expected number of years lost before age a* (Andersen 2013) as a basis for the comparison between countries.

In Section 2 we describe the data we use in our investigation, and we give a formula for the expected number of years lost before age *a*. In Section 3 we then compute the expected number of years lost for different cohorts *c* and ages *a*, and show how this makes it possible to obtain a fair comparison, also for younger cohorts, of how long women in Japan, Italy, Norway and Sweden may expect to live. In the final Section 4, we conclude and discuss our findings.

## 2. Data and Methods

We downloaded the period life tables with one-year age and calendar time intervals for Italy, Japan, Norway and Sweden from the Human Mortality Database (see list of references). From these life tables we extracted the one-year probabilities of death $q_{x,t}$, for all ages *x* and the periods $t = 1950, 1951, \ldots, 2012$, for each of the four countries.

A woman born in year *c* will be *x* years old in calendar year $t = c + x$. From the one-year probabilities of death $q_{x,t}$ from the period life tables, we may therefore obtain (an approximation to) the one-year probabilities of death for the cohort born in year *c* by $q_x^{(c)} = q_{x,c+x}$. These one-year probabilities of death may be combined in the usual way to obtain the expression $\ell_x^{(c)} = \prod_{i=0}^{x-1}\left(1 - q_i^{(c)}\right)$ for the probability that a woman who is born in year *c* will survive at least to age *x*. The expected number of years lived from birth to age *a* is then given by $\sum_{x=1}^{a} \ell_x^{(c)}$, and it follows that the expected number of years lost before age *a* for the cohort born in year *c* becomes $a - \sum_{x=1}^{a} \ell_x^{(c)}$.

## 3. Results

Figure 2 shows the expected number of years lost before age *a* as a function of *a* for the cohorts born in 1950, 1960, 1970 and 1980. For the cohorts born in 1950, we may compute the expected number of years lost before age *a* for ages up to $a = 62$ years. We see (Fig 2a) that for the 1950 cohort, the expected number of years lost are highest for Japanese women, but Italian women could expect to lose almost the same number of years. Sweden has the lowest expected number of years lost, but the number of lost years for Norwegian women are not much higher. At age 60, the expected number of years lost for Swedish and Norwegian women are 2.23 years and 2.52 years, respectively, while women in Japan and Italy could expect to lose more than twice as many years before age 60 (Japan: 5.73 years;

**Figure 2:** *Expected number of years lost before age a as a function of a for cohorts of women born in 1950, 1960, 1970 and 1980. Norway: red lines, Sweden: blue lines, Italy: green lines, Japan: black lines. Note that the scales are not the same for panels a-d.*

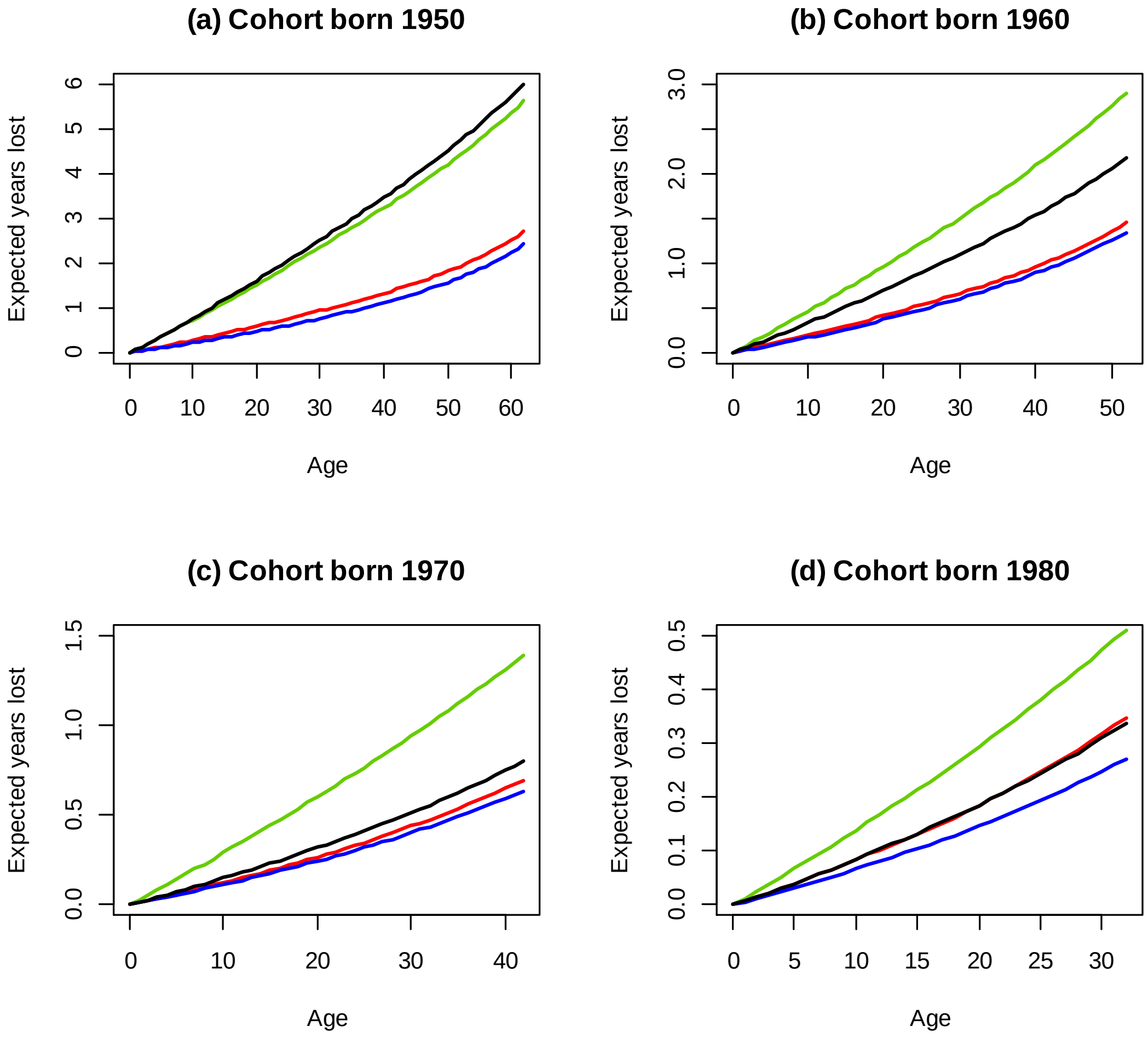

Italy: 5.36 years). For the cohorts born in 1960 and 1970, we may compute the expected number of years lost before age *a* for ages up to $a = 52$ and $a = 42$ years, respectively. We see (Figs 2b and 2c) that Sweden still has the lowest expected number of years lost, but the number of years lost for Norwegian women is only slightly higher. Also for the cohorts born in 1960 and 1970, Italian women could expect to lose more than twice as many years than the women from Sweden and Norway. But the situation for Japanese women is improved. For the cohort born in 1960 they are midway between the Scandinavian and

**Figure 3:** *Expected number of years lost before age 50 years, 40 years, 30 years, and 20 years as a function of birth cohort. Norway: red lines, Sweden: blue lines, Italy: green lines, Japan: black lines. Note that the scales are not the same for panels a-d.*

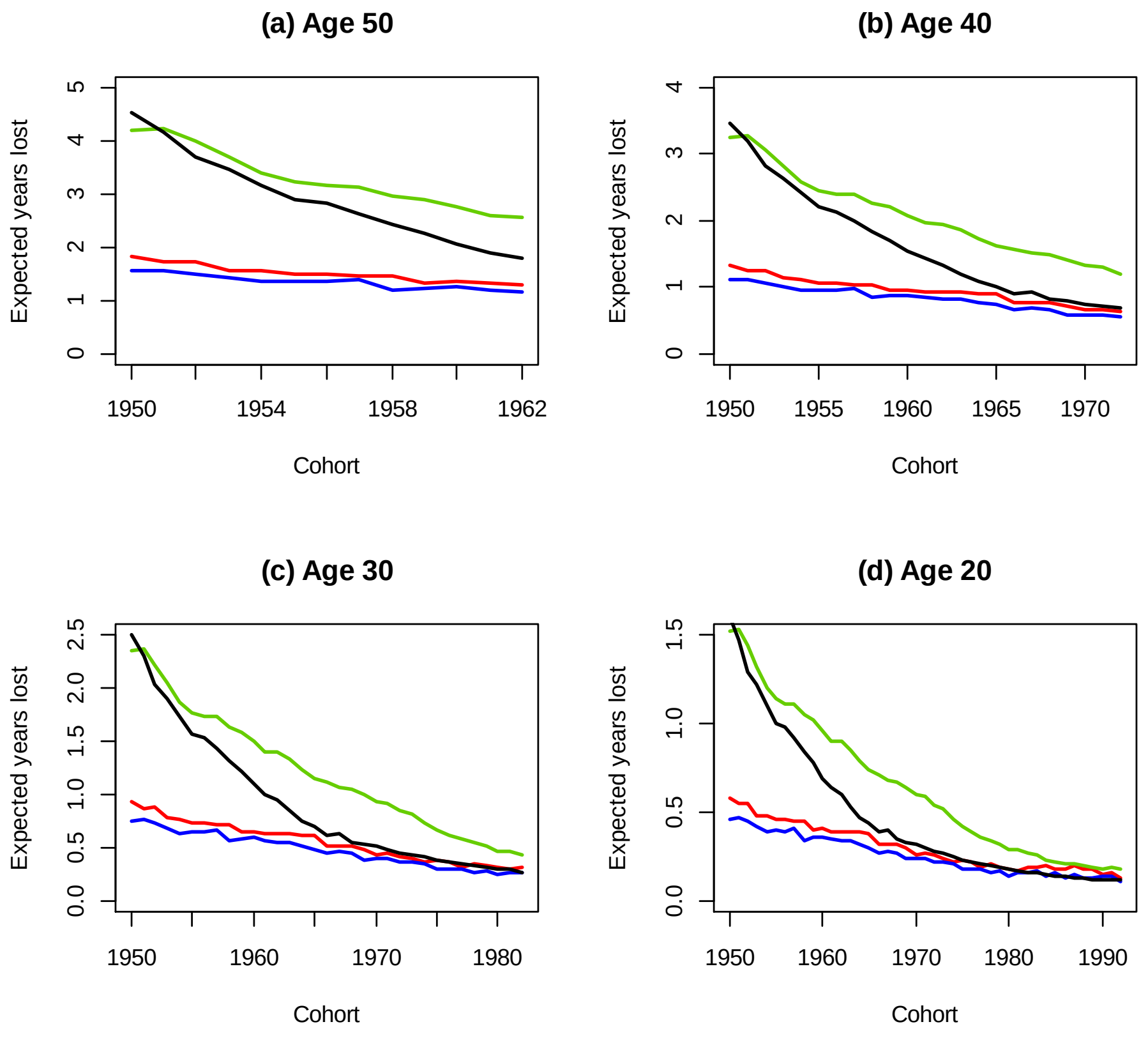

Italian women, and for the 1970 cohort the expected number of years lost for Japanese women are not much higher than for women from Sweden and Norway. Finally, we consider the cohorts born in 1980. We see (Fig. 2d) that also for these cohorts, Swedish women have the lowest expected number of years lost, while women in Italy may expect to lose about twice as many years. The number of years lost for Japanese and Norwegian women are almost indistinguishable and about 25 % higher than for Swedish women.

Figure 3 supplements the results of Figure 2. While we in Figure 2 show how the expected number of years lost before age *a* depends on *a* for four specified cohorts, we in Figure 3 show how the expected number of years lost before ages 50, 40, 30, and 20 years depend on the birth cohort. The figure shows that the expected number of years lost are reduced for the younger cohorts, and that the reduction is larger for women in Japan and Italy than for Scandinavian women (who already in 1950 could expect to lose quite few years). Further, for all the four ages considered and all cohorts (except the oldest ones),

Italian women could expect to lose more years than women in Japan and Scandinavia. For the oldest cohorts, the expected number of years lost for Japanese women are close to the number of years lost for women in Italy. But the number of years lost are reduced more quickly for Japanese women than for Italian women, and for the cohorts born in 1980 and later there are only minor differences between women in Japan and Scandinavia.

## 4. Discussion

If we consider life expectancy at birth based on period life tables (Fig 1), we get the impression that Italian women may expect to live longer than women in Scandinavia. However, from the cohort data (Figs 2 and 3) we see that the expected number of years lost for Italian women is substantially higher than for Swedish and Norwegian women, except for the cohorts born after 1990 where the number of years lost for Italy are only slightly larger than for Sweden and Norway. This shows that the period life expectancies are distorted by selection when they falsely give the impression that women in Italy may expect to live longer than Scandinavian women.

Japanese women have the longest life expectancy at birth in the world when computed from period data. In 2012, their expected life length was about 3 years longer than for Swedish and Norwegian women. But when we consider cohort data, we find that Japanese women born before 1980 may expect to lose more years than women in Scandinavia, and that there are only minor differences after 1980. Thus the cohort results provide no indication that Japanese women may expect to live longer than women in Scandinavia, so the large differences seen for period data may just be an artefact due to selection.

In this paper, we have focused on women in Italy, Japan, Norway, and Swden. The four countries may be classified into two vaguely defined groups: (i) countries where the living conditions improved earlier (Norway and Sweden), and (ii) countries where the improvement in living conditions occurred later (Italy and Japan). Our results are typical for what one will find when comparing one country from each of the two groups: If we consider life expectancies at birth based on period life tables, the country in group (i) will have the longest life expectancy back in time. But the life expectancy for the country in group (ii) will increase more rapidly, and may eventually become longer than the expected life length for the country in group (i). However, if we consider the expected number of years lost for different cohorts $c$ and ages $a$, it may be the case that people from the country in group (ii) may expect to lose more years than people from the country in group (i) for all cohorts $c$ and ages $a$. In such a situation, we cannot conclude that people in the country in group (ii) may expect to live longer than people in the country in group (i). This shows that differences in life expectancies based on period life tables may be caused by selection and not reflect real differences for the cohorts.